\documentstyle[preprint,aps,graphicx]{revtex}
\tightenlines

\begin{document}
\title{Ultracold collisions of oxygen molecules.}
\author{ Alexandr V. Avdeenkov and John L. Bohn}
\address{JILA and Department of Physics, University of Colorado, Boulder, CO}
\date{\today}
\maketitle

\begin{abstract}
  Collision cross sections and rate constants between two 
  ground- state oxygen molecules are
 investigated theoretically at translational energies below $\sim 1$K and in zero magnetic field.
 We present calculations for elastic and spin- changing inelastic collision rates
 for different isotopic combinations of oxygen atoms as a prelude to understanding
 their collisional stability in ultracold magnetic traps.
 A numerical  analysis has been made in the framework of a rigid- rotor model that 
 accounts fully for the singlet, triplet, and quintet  potential energy surfaces in this system. 
 The results offer insights into the effectiveness
 of evaporative cooling and the  properties of molecular Bose- Einstein condensates,
 as well as  estimates of collisional lifetimes in magnetic traps.
 Specifically, $^{17}O_{2}$ looks like a good candidate for ultracold studies, while
 $^{16}O_{2}$ is unlikely to survive evaporative cooling.
 Since $^{17}O_{2}$ is representative of a wide class of molecules that are paramagnetic in
 their ground state we conclude that many  molecules can be successfully magnetically trapped
 at ultralow temperatures.
\end{abstract}

\pacs{34.20.Cf, 34.50.-s, 05.30.Fk}

\narrowtext

\section{Introduction}
\subsection{Background}

Following the enormous successes of lowering the temperature of 
atoms to the sub-mK regime, experimental attention is now
turning to producing ultracold molecular samples.  The ``first
generation'' of cold molecule experiments has now demonstrated 
the efficient production of cold samples by a variety of techniques,
including photoassociation of ultracold atoms~\cite{fioretti}, counter-rotating
supersonic jets~\cite{gupta}, Stark slowing~\cite{meier}, and buffer-gas cooling~\cite{doyle}.  
The latter two have yielded trapped samples that are cold in
rotational, vibrational and translational degrees of freedom, although
translational temperatures are still in the $0.1$-$1$ K range.

The next generation of experiments will seek colder,
denser samples. One novel option for cooling molecules further is
the optical cooling strategy described in Ref. \cite{Vuletic}.
Alternatively, we consider in this paper the evaporative cooling
of paramagnetic molecules in a static magnetic trap, following 
the strategies that have been employed to produce ultracold
atoms.  The central issue to the success of this method is that the rate of
elastic, rethermalizing collisions far exceeds the rate of lossy inelastic 
collisions that produce untrapped, strong-field-seeking states.  In terms of
scattering rate coefficients, this criterion is usually written
$K_{\rm el} > 10^2 K_{\rm loss}$.
 A large ratio of $K_{\rm el}/K_{\rm loss}$ is also vital for the stability of the
 trapped gas once it is cold. A main objective of this paper is to demonstrate that 
 molecules with nonzero spin in their lowest energy state will be quite stable at ultralow temperatures.

We may reasonably assert that the elastic rate constants for neutral molecules
have roughly the same magnitudes as those for neutral atoms, $K_{\rm el}
\sim 10^{-12} - 10^{-10}$ cm$^3$/sec at low energies, barring unfortunately
placed zeros in their s-wave scattering cross sections. 
  Indeed, our calculations yield elastic rates of this
 magnitude.
 The spin- state- changing
rate constants are, however, completely unknown for molecules at ultralow
temperatures.  To rectify this situation the present paper presents pilot
calculations for cold collisions of molecular oxygen.  This work is 
a logical next step following Refs. \cite{bohn1,bohn2}, which considered
the interaction of molecular oxygen a the helium buffer gas.

Spin- state- changing cold collisions of O$_2$ molecules are driven by at least
three distinct physical processes, two of which are already familiar from
ultracold atom physics: i) Spin-exchange collisions, which typically lead
to unacceptably large loss rates for both atoms and molecules; ii) Spin-spin
magnetic dipolar interactions, which are typically small in either case; 
and iii) Spin-rotation interactions, unique to molecules, wherein electronic
spins are influenced by their coupling to rotational motion, which is in turn dependent 
on torques exerted by the anisotropic potential
energy surface(PES) between the molecules.  

Spin-exchange collisions can be avoided in cold molecule collisions, as
in cold atom collisions, by preparing the molecules in their ``stretched''
states, with mechanical rotation, electronic spin, and nuclear spin
(if any) all aligned along a common laboratory-fixed axis.  We will
therefore confine ourselves to this circumstance.  By far the leading
contribution to the rate constant for state- changing, lossy collisions~($K_{\rm loss}$)
 is then the spin-rotation coupling,
as shown below in detailed calculations.  Indeed, when an exothermic
exit channel is available, this coupling can yield loss rates comparable 
in magnitude to spin-exchange rates, i.e., comparable to elastic collision
rates.  This is the case for the $^{16}$O$_2$ molecule.  The 
spin- changing rate is, 
however, strongly suppressed when the only allowed exit channels are 
degenerate in energy with the incident channel and when the collision energy lies below 
a characteristic energy $E_0$.  In the stretched state of $^{17}$O$_2$
this is indeed the case, since changing the molecular spin at low energy requires
boosting the partial- wave angular momentum from $l=0$ to $l=2$. These collisions are
therefore suppressed by the Wigner threshold law when $E<E_{0}$,  where $E_{0}$ is the
height of the $l=2$ centrifugal barrier.
  For $^{17}$O$_2$ the barrier is 
roughly $E_0 \sim 0.013$ K, not far below the temperature that buffer-gas
cooling can take these molecules.

The potential for disaster in molecule cold collisions is far greater than
in atom cold collisions.  For example, hyperfine interactions are 
more complex, and include rotation-nuclear spin couplings that can invert
spins.  These are, however, expected to have minimal impact on the 
stretched-state molecules.
Also of potential significance are spin-vibration couplings, which we 
disregard in O$_2$ owing to the extremely large vibrational excitation
energy of ground state O$_2$ molecules compared to the energy available
to excite them.  The vibrational degrees of freedom
remain to be fully explored at ultralow temperatures, but it is expected
that vibrational quenching~(also an exothermic process) can occur with appreciable rates~\cite{Forrey}.  Finally,
polar molecules are susceptible to particularly strong long-range anisotropies.  While this
is not of direct relevance to molecular oxygen, it can be devastating to
the electrostatic trapping of polar molecules \cite{bohn3}, and potentially dangerous
for magnetically trapped molecules when the electric and molecular dipole
moments are coupled.

\subsection{Oxygen molecules- General considerations}

The importance of molecular oxygen as a potential candidate for
cooling and trapping experiments has been emphasized elsewhere 
\cite{bohn1,Friedrich}.  We will here consider O$_2$ molecules that have
been cooled  to temperatures
 below 1K.  We will furthermore assume that these molecules
have relaxed to their electronic  $^3\Sigma_g^-$ ground state, 
and $v=0$ vibrational ground state.  What remain are the rotational 
and spin degrees of freedom that influence evaporative cooling.  
In a typical magnetostatic trap the molecules can be confined 
provided that they are in a weak-field-seeking state, i.e.,
a state whose energy rises with magnetic field.  

The Zeeman diagram of O$_2$ is reproduced in Figure 1. Nuclear
exchange symmetry declares that homonuclear oxygen isotopomers
can have only even or only odd values of the nuclear rotation
quantum number $N$~\cite{mizushima}.  For $^{16}$O$_2$~(or $^{18}$O$_2$), 
which  has identical spin-zero nuclei, only odd values of $N$ are
allowed.  For the isotopomer $^{17}$O$_2$, where each nucleus has 
spin $I_N=5/2$, the allowed $N$ levels depend on the total nuclear
spin ${\vec I} = {\vec I_{NA}}  + {\vec I_{NB}}$.  The molecules
must have even values of $N$ for odd values of $I$, and vice-versa.
Thus the lowest energy weak-field seeking states are $|N,JM_J \rangle=
|1,22\rangle$ and $|1,21\rangle$ for $^{16}$O$_2$ or $^{18}$O$_2$,
and $|N,JM_J \rangle= |0,11\rangle$ for $^{17}$O$_2$. 
 These states are indicated
by heavy lines in the figure.

Figure 1 illustrates the essential difference between the even-$N$
and odd-$N$ manifolds in O$_2$, from the standpoint of inelastic collisions.
Namely, the trapped states with $J=2$ in the $N=1$ manifold can decay
exothermically to the untrapped $J=0$ states.  By contrast, the $N=0$
trapped state at low energy and low magnetic field can only change
its spin projection $M_J$ to other states that are nearly degenerate
in energy.  This difference proves crucial in strongly suppressing
spin-rotation collisions in $^{17}$O$_2$ relative to $^{16}$O$_2$.
This has already been discussed for cold collisions of O$_2$ with helium atoms
\cite{bohn1,bohn2}; the situation is similar when the molecules collide with each other.
As in Refs.~\cite{bohn1,bohn2}, we carry out calculations in zero magnetic field.

In this paper we ignore the role of nuclear spin, hence of hyperfine
structure, in the $^{17}$O$_2$ isotopomer.  This is justified
by considering the molecules to be in their stretched states of $N=0$,
with $M_J=J=1$ and $M_I=I=5$.  The nuclear spin degrees of freedom are
then frozen out.  In principle the weak nuclear
spin-rotation coupling would influence the nuclear spins, resulting
in $F=6 \rightarrow F=5,4$ transitions, but these couplings are an order- of-
magnitude smaller than the spin- rotation couplings we are already considering.  Moreover, measurements of
the $^{17}$O$_2$ microwave spectrum~\cite{Cazzoli} reveal that 
its hyperfine structure is inverted in its $N=0$ state, requiring that
$3.7mK$ of kinetic energy be supplied to change $F$.  
Thus at ultralow temperatures hyperfine- state- changing collisions are rigorously forbidden.

\section{Model}
\subsection{Hamiltonian}

Our investigation of molecular collisions follows the model of diatom-diatom
scattering originally due to~\cite{takay,heil}, but modified to incorporate the electronic spin of
oxygen molecules. The $O_2(^{3}\Sigma_{g}^{-}) - O_2(^{3}\Sigma_{g}^{-} )$ dimer has a 
spin-dependent intermolecular potential, namely three potential surfaces exist 
corresponding to singlet($S=0$), triplet($S=1$), and  quintet ($S=2$) states
of total electronic spin $S$~\cite{wormer}. 
The complete Hamiltonian for the collision process can be written
\begin{equation}
H = T_{A} + T_{B} + V_s + V_{lr} + V_{dd} + H_{rfs},
\end{equation}
where $T_i$ is the translational kinetic energy of molecule $i$; $V_s$
is the short-range exchange interaction; $V_{lr}$ is the long-range
potential consisting of dispersion and electric quadrupole-quadrupole
interactions; $V_{dd}$ is the electronic spin-spin dipolar interaction;
and $H_{rfs}$ is the Hamiltonian for the rotational fine structure of
the two separate
 oxygen molecules.  The hyperfine interaction will be disregarded for
now, as was discussed above.
The short-range potential can be written as
a mean interaction plus exchange corrections, following Ref.~\cite{avoird}:
\begin{equation}
\label{pes}
V_{s}({\bf R},\omega_{A},\omega_{B},{\bf S_{A}, S_{B}})=V_{av}({\bf R},\omega_{A},\omega_{B})
-2V_{ex}({\bf R},\omega_{A},\omega_{B}) {\bf S_{A}\cdot S_{B}},
\end{equation}
where $\omega=(\theta, \phi)$ are the polar angles of molecules A and B respectively, 
${\bf R}(R,\Theta)$ 
describes the radius vector between the center-of-mass of the molecules in the lab- fixed coordinate frame and
\begin{eqnarray}
\label{av}
V_{av}({\bf R},\omega_{A},\omega_{B}) =
\nonumber
\\ 
\sum_{L_{A},L_{B},L}
 f_{L_{A},L_{B},L}(R)
(-1)^{L_{B}-L_{A}}
\sqrt{(2L_{A}+1)(2L_{B}+1)} 
 {\bf K^{L}}(\omega_{A},\omega_{B}) {\bf \cdot C^{L}}(\Theta).
\end{eqnarray}
Here $K^{L}_{M} (\omega_{A},\omega_{B})= [C^{L_{A}}_{M_{A}} (\omega_{A})
 \otimes C^{L_{B}}_{M_{B}}(\omega_{B})]_{M}^{L}$ and $C_{M}^{L}$ are reduced spherical harmonics. 
 It should be said that the expansion~(\ref{av}) is identical to that in~\cite{avoird} but written
 in terms of reduced spherical harmonics. The spin-dependent Heisenberg exchange
term $V_{ex}$ is expanded similarly, but with a different expansion coefficient $g_{L_{A},L_{B},L}(R)$.
The expressions for expansion coefficients $f_{L_{A},L_{B},L}(R)$ and $g_{L_{A},L_{B},L}(R)$ were obtained in the
work of~\cite{wormer}, and the quadrupole- quadrupole interaction has a similar form. 
The $C_{6}$ dispersion coefficients were calculated in~\cite{bussery} in the body- fixed frame.
To unify our treatment we recast the anisotropic $C_6$ coefficients in terms of 
the same angular basis as the exchange potential:
\begin{eqnarray}
\label{disp}
V_{disp}({\bf R},\omega_{A},\omega_{B}) =
\nonumber
\\ 
- \sum_{L_{A},L_{B},L}
 \frac{p_{L_{A},L_{B},L}}{R^{6}}
(-1)^{L_{B}-L_{A}}
\sqrt{(2L_{A}+1)(2L_{B}+1)}
{\bf K^{L}}(\omega_{A},\omega_{B}) {\bf \cdot C^{L}}(\Theta),
\end{eqnarray}
where
\begin{equation}
\label{coef}
p_{L_{A},L_{B},L}= \sqrt{\frac{2L+1}{(2L_{A}+1)(2L_{B}+1)}} \sum_{M_{A},M_{B}} d_{L_{A},L_{B},M_{A},M_{B}}
\left( \begin{array}{ccc}
                  L_{A} & L_{B} & L \\
                  M_{A} & M_{B} & 0
                 \end{array} \right)
\end{equation}
The connection between our coefficients $d_{L_{A},L_{B},M_{A},M_{B}}$ and coefficients from~\cite{bussery} is
in Appendix A.
Figure 2  shows a slice through the potential energy surface for the singlet 
states of the $[O_2(^{3}\Sigma_{g}^{-})]_2$ dimer for the ``H''-geometry of the 
two molecules, which passes through the global minimum of the PES. 
Also shown are contributions to this potential from various components 
with different $(L,L_{A},L_{B})$.  We can see that this potential has a
very strong anisotropy.  Notice that the isotropic contribution,
with $L_A=L_B=L=0$ accounts for less than half of the total well depth. 

As in the case of ultracold atoms, details of ultracold molecular collisions
depend extremely sensitively on details of PES. Thus eventually the PES must be
fine- tuned using key experimental data to provide complete quantitative results~\cite{Hudson,Avdeenkov}.
Nevertheless, the order-of- magnitude of the rate constants, and their general trends, already emerge
clearly in the present model.

The intermolecular spin-spin (magnetic dipole) interaction has the form~\cite{avoird}:
\begin{eqnarray}
\label{dipol}
V_{dd}({\bf R, S_{A},S_{B}})= -\frac{\sqrt{6}g_{e}^{2}\mu_{B}^{2}}{R^{3}}
{\bf C^{2}}(\Theta){\bf \cdot}[{\bf S_{A}\otimes S_{B}}]^{\bf 2},
\end{eqnarray}
where ${\bf R}\equiv (R, \Theta)$, $g_{e}=2.0023$ and $\mu_{B}$ is the Bohr magneton.
For alkali atoms this is known to be a fairly weak contribution to
spin-changing collisions, a conclusion that we find holds for molecules as well.
Finally, in our model we take into account the molecular fine structure, which
arises  from the molecular rotation and spin-rotation coupling and is diagonal in our
total-spin basis at large R. For $^{16}O_{2}$ we use the fine- structure constants 
determined in~\cite{tinkham}, and for $^{17}O_{2}$ we employ those 
determined in~\cite{gazzoli}.

We express the Hamiltonian in a basis of total angular momentum,
\begin{eqnarray}
\label{basis}
\nonumber
{\cal J} = {\bf J} + {\bf l}
\\
{\bf J} = {\bf J_{1}} + {\bf J_{2}}
\\
\nonumber
{\bf J_{i}} =  {\bf S_{i}} + {\bf N_{i}},
\end{eqnarray}
in the terms of each molecule's mechanical rotation (${\bf N_{i}}$), its electronic spin
  (${\bf S_{i}}$), its total spin (${\bf J_{i}}$), the combined spin for
  two molecules together (${\bf J}$), and the partial wave representing the rotation of the molecule about
the  center of mass (${\bf l}$).  In zero magnetic field both ${\cal J}$ and
its laboratory-fixed projection ${\cal M}$ are rigorously conserved.
  
In this basis we can present our wave function as:
\begin{eqnarray}
\label{wf}
\nonumber
\Psi_{\cal J,M}(R,\Theta,\omega_{A},\omega_{B},\sigma_{A}, \sigma_{B})=
 \frac{1}{R} \sum_{l,{\cal J},J_{1},J_{2},N_{1},N_{2},S_{1},S_{2}}
\psi_{l,{\cal J},J_{1},J_{2},N_{1},N_{2},S_{1},S_{2}}(R)
\\
*I_{{\cal J,M};l, J ,J_{1},J_{2},N_{1},N_{2},S_{1},S_{2}}(\Theta,\omega_{A},\omega_{B},\sigma_{A}, \sigma_{B}),
\end{eqnarray}
where $\sigma_{A,B}$ represent the electronic spin coordinates for molecule A,B.
The coupled angular momentum basis functions are defined by suitable
tensor products:
\begin{eqnarray}
\label{angwf}
I_{{\cal J,M};l, J ,J_{1},J_{2},N_{1},N_{2},S_{1},S_{2}}(\Theta,\omega_{A},\omega_{B},\sigma_{A}, \sigma_{B})=
[P^{J}_{M_{J}}(\omega_{A},\omega_{B},\sigma_{A}, \sigma_{B}) \otimes Y^{l}_{m_{l}}(\Theta)]^{\cal J}_{\cal M},
\end{eqnarray}
\begin{eqnarray}
P^{J}_{M_{J}}(\omega_{A},\omega_{B},\sigma_{A}, \sigma_{B}) =[T^{J_{1}}_{M_{J_{1}}}(\omega_{A}, \sigma_{A})
\otimes T^{J_{2}}_{M_{J_{2}}}(\omega_{B}, \sigma_{B})]^{J}_{M_{J}},
\end{eqnarray}
\begin{eqnarray}
T^{J_{i}}_{M_{J_{i}}}(\omega, \sigma) = [Y^{N_{i}}_{M_{N_{i}}}(\omega) \otimes \chi^{S_{i}}_{M_{S_{i}}}
(\sigma)]^{J_{i}}_{M_{J_{i}}},
\end{eqnarray}
where Y is a spherical harmonic and $\chi$ is a spinor wave function.

Because   target and projectile are identical bosons, we must take into account the symmetry of the
wave function under exchange. To this end we construct symmetrized 
angular momentum functions from~(\ref{angwf}
):
\begin{eqnarray}
\label{sangwf}
I_{l,J,J_{1},J_{2},N_{1},N_{2},S_{1},S_{2}}^{s}=\frac{I_{l,J,J_{1},J_{2},N_{1},N_{2},S_{1},S_{2}} +
 (-1)^{J+J_{1}+J_{2}+l} I_{l,J,J_{2},J_{1},N_{2},N_{1},S_{2},S_{1}}}
{\sqrt{2(1+\delta_{J_{1}J_{2}} \delta_{N_{1}N_{2}} \delta_{S_{1}S_{2}})}}
\end{eqnarray}
 We have omitted the conserved quantum numbers $\cal J,M $ in this expression.  

 To calculate the reduced matrix elements in our basis we  recouple
the angular part of the potential in terms of tensor elements:
\begin{eqnarray}
\label{tenz}
K_{M}^{L}(\omega_{A},\omega_{B})[{\bf S_{A}\cdot S_{B}}]^{\gamma}= \sum_{J_{A},J_{B}} \sqrt{[J_{A}][J_{B}]}
(-1)^{L-L_{A}-J_{B}}
\left\{ \begin{array}{ccc}
                  L_{A} & L_{B} & \gamma \\
                  J_{B} & J_{A} & L
                 \end{array} \right\}
\\
\nonumber
\times
[T^{J_{A}}_{M_{J_{A}}} \otimes T^{J_{B}}_{M_{J_{B}}}]^{L}_{M_{L}},
\end{eqnarray}
in terms of the spherical tensors
$T^{J_{A}}_{M_{J_{A}}}(\omega_{A},\sigma_{A}) = [Y^{L_{A}}_{M_{L_{A}}} \otimes [\chi^{S_{A}
}_{M_{S_{A}}}]^{\gamma}]^{J_{A}}_{M_{J_{A}}}$, with 
$\gamma=0$ for $V_{av}$ and $\gamma=1$  for the exchange part of the 
interaction~(\ref{pes}), and $[Q]=(2Q+1)$. 

Using our expansion of the inter- molecular potential~(\ref{av}) and~(\ref{disp}), the wave function~(\ref{wf})
and taking into account the Wigner - Eckart theorem
,
\begin{eqnarray}
<{\cal J M}| V_{s}+V_{disp}+V_{dd}|{\cal J^{\prime} M^{\prime}}>=\delta_{\cal J J^{\prime}} \delta_{\cal M M^{\prime}}
<{\cal J }|| V_{s}+V_{disp}+V_{dd}||{\cal J^{\prime}}>
,
\end{eqnarray}
 we can present the reduced matrix 
elements for the $V_{s}+V_{lr}$ part as:
\begin{eqnarray}
\label{reduc}
\nonumber
<  \{[J_{1}(N_{1}S_{1})J_{2}(N_{2}S_{2})]Jl\}{\cal J }||
({\bf  K^{L} \cdot {C^{L}}})
[{\bf S_{A}\cdot S_{B}}]^{\gamma}
|| \{[J_{1}^{\prime}(N_{1}^{\prime}S_{1}^{\prime})J_{2}^{\prime}(N_{2}^{\prime}S_{2}^{\prime})]J^{\prime}l^{\prime}\}{\cal J }>
\nonumber
\\
=(-1)^{{\cal J}+J^{\prime}-L_{A}+N_{1}+N_{2}}([l^{\prime}][J^{\prime}][l][J][J_{1}^{\prime}][J_{2}^{\prime}]
[J_{1}][J_{2}][N_{1}^{\prime}][N_{2}^{\prime}][N_{1}][N_{2}])^{1/2}
\nonumber
\\
\times
\left\{ \begin{array}{ccc}
                  l & l^{\prime} & L \\
                  J^{\prime} & J & {\cal J}
                 \end{array} \right\}                
\left( \begin{array}{ccc}
                  l & L & l^{\prime} \\
                  0 & 0 & 0
                 \end{array} \right)
\left( \begin{array}{ccc}
                  N_{1} & L_{A} & N_{1}^{\prime} \\
                  0 & 0 & 0
                 \end{array} \right)
\left( \begin{array}{ccc}
                  N_{2} & L_{B} & N_{2}^{\prime} \\
                  0 & 0 & 0
                  \end{array} \right)
\\                  
\nonumber
\times
\sum_{J_{A}J_{B}} (-1)^{J_{B}}[J_{A}][J_{B}] 
\left\{ \begin{array}{ccc}
                  L_{A} & L_{B} & \gamma \\
                  J_{B} & J_{A} & L
                 \end{array} \right\}
\left\{ \begin{array}{ccc}
                  J & J^{\prime} & L \\
                  J_{1} & J_{1}^{\prime} & J_{A} \\
                  J_{2} & J_{2}^{\prime} & J_{B}
                 \end{array} \right\} 
\left\{ \begin{array}{ccc}
                  J_{1} & J_{1}^{\prime} & J_{A} \\
                  N_{1} & N_{1}^{\prime} & L_{A} \\
                  S_{1} & S_{1}^{\prime} & \gamma
                 \end{array} \right\} 
\left\{ \begin{array}{ccc}
                  J_{2} & J_{2}^{\prime} & J_{B} \\
                  N_{2} & N_{2}^{\prime} & L_{B} \\
                  S_{2} & S_{2}^{\prime} & \gamma
                 \end{array} \right\} 
\\
\nonumber                 
\times
\Bigl( \sqrt{S_{A}(S_{A}+1)(2S_{A}+1)S_{B}(S_{B}+1)(2S_{B}+1)} \Bigr)^{\gamma}                 
\delta_{S_{1}S_{1}^{\prime}} \delta_{S_{2}S_{2}^{\prime}},                                                               
\end{eqnarray}
and for the $V_{dd}$ part as:
\begin{eqnarray}
\nonumber
<  \{[J_{1}(N_{1}S_{1})J_{2}(N_{2}S_{2})]Jl\}{\cal J }||
\bf{(C^{2} \cdot [S_{A} \otimes S_{B}]^{2})}
|| \{[J_{1}^{\prime}(N_{1}^{\prime}S_{1}^{\prime})J_{2}^{\prime}(N_{2}^{\prime}S_{2}^{\prime})]J^{\prime}l^{\prime}\}{\cal J }>
\\
=(-1)^{{\cal J}+J^{\prime}+N_{1}+N_{2}+J_{1}+J_{2}}([l^{\prime}][J^{\prime}][l][J][J_{1}^{\prime}][J_{2}^{\prime}]
[J_{1}][J_{2}][2])^{1/2}
\left\{ \begin{array}{ccc}
                  l & l^{\prime} & 2 \\
                  J^{\prime} & J & {\cal J}
                 \end{array} \right\}
\\                 
\nonumber                
\times
\left( \begin{array}{ccc}
                  l & 2 & l^{\prime} \\
                  0 & 0 & 0
                 \end{array} \right)
\left\{ \begin{array}{ccc}
                  J_{1} & J_{1}^{\prime} & 1 \\
                  1 & 1 & N_{1}
                 \end{array} \right\}
\left\{ \begin{array}{ccc}
                  J_{2} & J_{2}^{\prime} & 1 \\
                  1 & 1 & N_{2}
                 \end{array} \right\}
\left\{ \begin{array}{ccc}
                  J & J^{\prime} & 2 \\
                  J_{1} & J_{2} & 1 \\
                  J_{1}^{\prime} & J_{2}^{\prime} & 1
                 \end{array} \right\} 
\\
\nonumber  
\times               
\sqrt{S_{A}(S_{A}+1)(2S_{A}+1)S_{B}(S_{B}+1)(2S_{B}+1)}  
\delta_{S_{1}S_{1}^{\prime}} \delta_{S_{2}S_{2}^{\prime}}
\delta_{N_{1}N_{1}^{\prime}} \delta_{N_{2}N_{2}^{\prime}},   
\end{eqnarray}

The reduced matrix elements of our potential between the states defined by~(\ref{wf}) and using~(\ref{sangwf}) are
\begin{eqnarray}
\label{sreduc}
\nonumber
< \eta J_{1} J_{2}|| U^{s} || \eta^{\prime} J_{1}^{\prime} J_{2}^{\prime}>=
\frac{< \eta J_{1} J_{2}|| U || \eta^{\prime} J_{1}^{\prime} J_{2}^{\prime}>
+ (-1)^{J+J_{1}+J_{2}+l}
< \eta J_{2} J_{1}|| U || \eta^{\prime} J_{1}^{\prime} J_{2}^{\prime}>}
{\sqrt{(1+\delta_{J_{1}J_{2}} \delta_{N_{1}N_{2}}) (1+\delta_{J_{1}^{\prime}J_{2}^{\prime}}\delta_{N_{1}^{\prime}N_{2}^{\prime}})}}
\\
\times
 \frac{1+(-1)^{N_{1}+N_{2}+l+N_{1}^{\prime}+N_{2}^{\prime}+l^{\prime}}}{2},
\end{eqnarray}
where $\eta$ stands for the rest of the quantum numbers from our wave function~(\ref{wf}). The coupling matrix element therefore vanishes between channels with 
different  parity $(-1)^{N_{1}+N_{2}+l}$.

Figure 3 shows a partial set of adiabatic potential curves for $^{16}O_{2}$ in the case of ${\cal J}=0$. 
To generate this figure we include rotational channels $N=1,3,5$ and even partial waves $l=0-6$, which
already imply 100 channels in this case. The strong anisotropy in the PES is here manifested mainly
in a set of strongly avoided crossings near $R=8$ a.u.  For $R>8$ a.u. the PES strongly favors
a collinear configuration of the pair of molecules, while for $R<8$ a.u. it strongly favors
a parallel, ``H''-shaped configuration.  
This figure stresses the importance of higher-lying rotational states in determining the
details of scattering even at ultracold energies.  However,
for $^{16}O_{2}$ we have chosen to compute cross sections just for the case
$l=0-10, N=1$ because these calculations already reveal very large spin-changing rates.
Higher-lying channels will influence the details, but are unlikely to suppress losses.
The total number of channels, including all values of $\cal J$, is then 212.
For $^{17}O_{2}$, by contrast, we have computed cross sections for $l=0- 10$ and $N=0,2$,
to verify that higher-lying rotational states do not upset the observed suppression of loss
rates.  In this case the total number of channels considered is therefore 836.

\subsection{Evaluating cross sections}
We solve the coupled-channel equations using a log-derivative propagator method \cite{Johnson}
to determine scattering matrices.  Since we assume zero magnetic field the total
angular momentum ${\cal J}$ is a good quantum number and the results are independent of the laboratory
projection ${\cal M}$ of total angular momentum.

For magnetic trapping the molecular quantum numbers of interest are naturally the magnetic
quantum numbers.  Therefore we need to know the state- to- state cross sections in the
$|N_{1}N_{2}J_{1}J_{2},M_{J_{1}}M_{J_{2}}>$ basis.  The scattering matrices are readily converted to this basis: 
\begin{eqnarray}
\label{sm}
\nonumber
<N_{1} N_{2} J_{1} J_{2} M_{J_{1}}  M_{J_{2}} l M_{l}| {\cal S} |
N_{1}^{\prime} N_{2}^{\prime} J_{1}^{\prime} J_{2}^{\prime} M_{J_{1}}^{\prime}  M_{J_{2}}^{\prime}
l^{\prime} M_{l}^{\prime}> =
\\
\nonumber
\sum_{J J^{\prime}} <J_{1} M_{J_{1}} J_{2} M_{J_{2}}| J M_{J}> 
<J^{\prime} M_{J}^{\prime}|J_{1}^{\prime} M_{J_{1}}^{\prime} J_{2}^{\prime} M_{J_{2}}^{\prime}>
\\
\times \sum_{\cal J} <J M_{J} l M_{l} |{\cal J M_{J}}> <{\cal J M_{J}} | J^{\prime} M_{J}^{\prime} l^{\prime} M_{l}^{\prime}>
\\
\nonumber
\times
<  \{[J_{1}(N_{1}S_{1})J_{2}(N_{2}S_{2})]Jl\}{\cal JM }|
{\cal S(J)}
| \{[J_{1}^{\prime}(N_{1}^{\prime}S_{1}^{\prime})J_{2}^{\prime}(N_{2}^{\prime}S_{2}^{\prime})]J^{\prime}l^{\prime}\}{\cal JM }>
\end{eqnarray}

For notational simplicity we define the index $\alpha = (N_{1} N_{2} J_{1} J_{2})$ in the following.
Then the complete symmetrized wave function in the limit of large R is given in~\cite{takay}:
\begin{eqnarray}
\label{scat1}
\nonumber
\frac{(\exp (i \vec{k}_{\alpha} \cdot \vec{R}) 
T^{J_{1}}_{M_{J_{1}}}(\omega_{A}, \sigma_{A}) 
T^{J_{2}}_{M_{J_{2}}}(\omega_{B}, \sigma_{B}) +
\exp (- i \vec{k}_{\alpha} \cdot \vec{R}) 
T^{J_{1}}_{M_{J_{1}}}(\omega_{B}, \sigma_{B}) 
T^{J_{2}}_{M_{J_{2}}}(\omega_{A}, \sigma_{A}) }{\sqrt{2}}
\\
+ \sum_{\alpha^{\prime} M_{J_{1}}^{\prime} M_{J_{2}}^{\prime} }
\frac{\exp (i k_{\alpha}^{\prime}  R)}{R} \cdot
\frac{f_{J_{1}^{\prime}M_{J_{1}}^{\prime} J_{2}^{\prime}M_{J_{2}}^{\prime}}(\hat{R})
+
f_{J_{2}^{\prime}M_{J_{2}}^{\prime} J_{1}^{\prime}M_{J_{1}}^{\prime}}(- \hat{R}) 
}{\sqrt{2}}
T^{J^{\prime}_{1}}_{M_{J^{\prime}_{1}}}(\omega_{A}, \sigma_{A}) 
T^{J^{\prime}_{2}}_{M^{\prime}_{J_{2}}}(\omega_{B}, \sigma_{B})
\end{eqnarray}
where $f_{J_{1}^{\prime}M_{J_{1}}^{\prime} J_{2}^{\prime}M_{J_{2}}^{\prime}}$ is the channel-dependent
scattering amplitude.

Using the definition of our wave function~(\ref{wf}) and transforming it into the $|J_{1}J_2M_{J_{1}}M_{J_{2}}>$ basis we
can get  the asymptotic form of the wave function in terms of the  $S$- matrix~\cite{takay}:
\begin{eqnarray}
\label{scat2}
\nonumber
\sum_{{\cal J}, J, l, M_{l}} <J_{1} M_{J_{1}} J_{2} M_{J_{2}}| J M_{J}> <J M_{J} l M_{l} |{\cal J M_{J}}>
\sqrt{1+\delta_{J_{1}J_{2}} \delta_{N_{1}N_{2}} \delta_{M_{J_{1}}M_{J_{2}} }}
\\
\times
\frac{4\pi}{2i \sqrt{k_{\alpha}}R} i^{l} Y^{l}_{M_{l}}(\hat{k}_{\alpha}) 
\sum_{l^{\prime}, M^{\prime}_{l}}
\frac{1}{ \sqrt{k^{\prime}_{\alpha}}} 
 \sum_{J^{\prime},J_{1}^{\prime},J_{2}^{\prime}}
(\delta_{J_{1}J_{1}^{\prime}} \delta_{J_{2}J_{2}^{\prime}} \delta_{J J^{\prime}} \delta_{l l^{\prime}}
\exp (-i (k_{\alpha}^{\prime} \cdot R - l^{\prime}\pi /2)
\\
\nonumber
-\exp (i (k_{\alpha}^{\prime} \cdot R - l^{\prime}\pi /2)
<\{[J_{1}J_{2}]Jl\} {\cal J M}|{\cal S(J)}|\{[J_{1}^{\prime}J_{2}^{\prime}]J^{\prime} l^{\prime}\} {\cal J M} >)
\\
\nonumber
\times
I^{s}_{{\cal J,M};l^{\prime}, J^{\prime} ,J_{1}^{\prime},J_{2}^{\prime}}(\Theta,\omega_{A},\omega_{B},\sigma_{A}, \sigma_{B}),
\end{eqnarray}

By comparing~(\ref{scat1}) with~(\ref{scat2}), we obtain the expression for the scattering amplitude:
\begin{eqnarray}
\label{ampl}
\nonumber
f_{J_{1}^{\prime}M_{J_{1}}^{\prime} J_{2}^{\prime}M_{J_{2}}^{\prime}}(\hat{R})
+
f_{J_{2}^{\prime}M_{J_{2}}^{\prime} J_{1}^{\prime}M_{J_{1}}^{\prime}}(- \hat{R}) 
=\frac{4\pi}{2i \sqrt{k_{\alpha} k_{\alpha}^{\prime}}}
\\
\times
\sum_{{\cal M}, l, l^{\prime}} \sqrt{1+\delta_{J_{1}J_{2}}\delta_{N_{1}N_{2}}\delta_{M_{J_{1}}M_{J_{2}}} } 
\sqrt{1+\delta_{J_{1}^{\prime}J_{2}^{\prime}}\delta_{N_{1}^{\prime}N_{2}^{\prime}}\delta_{M_{J_{1}}^{\prime} M_{J_{2}}^{\prime}}  }
Y^{l}_{M_{l}}(\hat{k_{\alpha}})i^{l- l^{\prime}}
\\
\nonumber
\times
<N_{1} N_{2} J_{1} M_{J_{1}} J_{2} M_{J_{2}} l M_{l}| {\cal S -  I} |
N_{1}^{\prime} N_{2}^{\prime} J_{1}^{\prime} M_{J_{1}}^{\prime} J_{2}^{\prime} M_{J_{2}}^{\prime}
l^{\prime} M_{l}^{\prime}> 
Y^{l^{\prime}}_{M_{l}^{\prime}}(\hat{R}),
\end{eqnarray}
where
 in the symmetrized separate- molecule basis the scattering matrix is given by
\begin{eqnarray}
\label{smsym}
\nonumber
<N_{1} N_{2} J_{1} J_{2} M_{J_{1}} M_{J_{2}} l M_{l}| {\cal S} |
N_{1}^{\prime} N_{2}^{\prime} J_{1}^{\prime} J_{2}^{\prime} M_{J_{1}}^{\prime} M_{J_{2}}^{\prime}
l^{\prime} M_{l}^{\prime}> =
\\
\nonumber
\frac{\sqrt{1+\delta_{J_{1}J_{2}}\delta_{N_{1}N_{2}}}}
{\sqrt{1+\delta_{J_{1}J_{2}}\delta_{N_{1}N_{2}}\delta_{M_{J_{1}}M_{J_{2}}} }}
\sum_{J J^{\prime}} <J_{1} M_{J_{1}} J_{2} M_{J_{2}}| J M_{J}> 
<J^{\prime} M_{J}^{\prime}|J_{1}^{\prime} M_{J_{1}}^{\prime} J_{2}^{\prime} M_{J_{2}}^{\prime}>
\\
\times 
\frac
{\sqrt{1+\delta_{J_{1}^{\prime}J_{2}^{\prime}}\delta_{N_{1}^{\prime}N_{2}^{\prime}} }}
{\sqrt{1+\delta_{J_{1}^{\prime}J_{2}^{\prime}}\delta_{N_{1}^{\prime}N_{2}^{\prime}}\delta_{M_{J_{1}}^{\prime} M_{J_{2}}^{\prime}}  }}
\sum_{\cal J} <J M_{J} l M_{l} |{\cal J M_{J}}> <{\cal J M_{J}} | J^{\prime} M_{J}^{\prime} l^{\prime} M_{l}^{\prime}>
\\
\nonumber
\times
<  \{[J_{1}(N_{1}S_{1})J_{2}(N_{2}S_{2})]Jl\}{\cal JM }|
{\cal S(J)}
| \{[J_{1}^{\prime}(N_{1}^{\prime}S_{1}^{\prime})J_{2}^{\prime}(N_{2}^{\prime}S_{2}^{\prime})]J^{\prime}l^{\prime}\}{\cal JM }>.
\end{eqnarray}
Symmetrization in the $|J_1J_2M_{J_1}M_{J_2} \rangle$ basis requires that
$J_1 \ge J_2$ and that $M_1 \ge M_2$ when $J_1=J_2$.

To obtain the scattering cross section we must integrate over the angular coordinates of the scattered wave.
But, for undistinguishable final spin states we  restrict the integral over half space ($\int d \Theta =
2 \pi$) to avoid double counting~\cite{burke}.

The total state- to- state cross section of interest for spin-rotational excitation and relaxation phenomena
can be obtained using~(\ref{ampl}) from the $\cal S$ matrix:
\begin{eqnarray}
\label{cross}
\nonumber
\sigma_{(N_{1} N_{2}) J_{1} J_{2} M_{J_{1}} M_{J_{2}} \rightarrow 
(N_{1}^{\prime} N_{2}^{\prime}) J_{1}^{\prime} J_{2}^{\prime} M_{J_{1}}^{\prime} M_{J_{2}}^{\prime}}=
\frac{(1+\delta_{J_{1}J_{2}}\delta_{N_{1}N_{2}}\delta_{M_{J_{1}}M_{J_{2}}}) \pi}{k^{2}_{N_{1} N_{2} J_{1} J_{2}}} 
\\ \times
\sum_{l M_{l} l^{\prime} M_{l}^{\prime}}
|<(N_{1} N_{2}) J_{1} M_{J_{1}} J_{2} M_{J_{2}} l M_{l}| {\cal S -  I} |
(N_{1}^{\prime} N_{2}^{\prime}) J_{1}^{\prime} M_{J_{1}}^{\prime} J_{2}^{\prime} M_{J_{2}}^{\prime}
l^{\prime} M_{l}^{\prime}> |^{2},
\end{eqnarray}
where 
\begin{eqnarray}
k_{N_{1} N_{2} J_{1} J_{2} } =(2\mu (E-E_{N_{1} J_{1}}- E_{N_{2} J_{2}}))^{1/2}
\end{eqnarray}
is the channel wavenumber and $E_{N_{1,2} J_{1,2}}$ are fine structure energy levels.
In this expression we assume an average over all incident directions, as in \cite{bohn1}.
Finally, state- to- state rate coefficients are given by
\begin{eqnarray}
\label{rate}
K_{(N_{1} N_{2}) J_{1} J_{2} M_{J_{1}} M_{J_{2}} \rightarrow 
(N_{1}^{\prime} N_{2}^{\prime}) J_{1}^{\prime} J_{2}^{\prime} M_{J_{1}}^{\prime} M_{J_{2}}^{\prime}}=
\upsilon_{(N_{1} N_{2}) J_{1} J_{2}  } 
\sigma_{(N_{1} N_{2}) J_{1} J_{2} M_{J_{1}} M_{J_{2}} \rightarrow 
(N_{1}^{\prime} N_{2}^{\prime}) J_{1}^{\prime} J_{2}^{\prime} M_{J_{1}}^{\prime} M_{J_{2}}^{\prime}},
\end{eqnarray}
where $\upsilon_{(N_{1} N_{2}) J_{1} J_{2}  }$ is the relative velocity of the collision
partners before the collision.

\section{Results}
This paper considers the scattering problem  for the homonuclear species $^{16}O_{2}$ and $^{17}O_{2}$.
 Ref.~\cite{bohn1,bohn2} speculated that buffer- gas cooling of ${O}_{2}$ by Helium should be possible,
  thus lowering the molecules to typical temperatures $\approx 0.3K$
   To further cool the gas by evaporative cooling requires favorabale collision rates for collision energies
  $E \lesssim 1K$. We will limit our detailed calculations to this case.
We will see that the cooling of $^{17}O_{2}$ could be quite efficient, while it is probably impossible for $^{16}O_{2}$.

\subsection{$^{17}O_{2}$- elastic scattering.}
Since  $^{17}O_{2}$ is the most promising candidate for evaporative cooling, we devote our attention to 
this isotopomer.
We focus on the
  $|N_{1}N_{2}J_{1}J_{2},M_{J_{1}} M_{J_{2}}> = |0011,11>$ state which is the lowest- lying trappable state
 for the even N- manifold~(Fig.1). 
 
 For identical bosons only even $l$-  partial waves contribute to the  cross sections for the   $|0011,11>$ state.
 A first important point of our calculations is to determine the number of partial waves that contribute to the
 cross section in the energy region up to 1K and how many molecular rotational levels should be taken into account.
 In principle  many partial waves are coupled together by the very anisotropic potential, but higher
 partial waves are suppressed at a low energy.
 Figure 4 illustrates the elastic cross sections for  different valves of the highest partial wave included. 
This Figure shows that it is enough to include just l=0-4 partial waves for the qualitative description of the cross section in
the region up to $\approx 0.2K$, and that the partial waves
l=0-10 are sufficient in the region up to $\approx 1K$.
 For all our calculations for $^{17}O_{2}$ we considered just the two lowest rotational levels $N=0,2$. 
 Including only the
  $N=0$ rotational level allows the molecules to explore only the
   isotropic part of their PES. Thus $N=2$ states must be included at least.
   The influence of higher rotational levels are found to be small in test calculations,  
   although they  impact details of the resonance structure.
    The calculations thus include 836 channels.
    
Particularly striking in Figure 4 is the strong difference in the 
cross sections when $l_{max}=2$ as opposed to $l_{max}=4,6,8,10$.
This is caused by the strong
anisotropy of the potential~\cite{wormer} and mathematically this means that for the case $l_{max}=2$ we
take into account only a small number of expansion functions~$(L, L_{A}, L_{B})$ in~(\ref{av}). 
Although, these few functions represent "most" of the potential, 
even small changes in potential can change the behavior of the cross section dramatically near zero energy~\cite{Avdeenkov}.
 Figure 5 shows the elastic cross sections for  $|0011,11>$ collisions and the
contribution from different partial waves. We can see that the partial waves $l=8$ and 
$l=10$  contribute significantly only above $\approx 0.5K$.

Figure 5 also exhibits dozens  of resonances  for molecular- molecular collisions below 1K, arising from
 the enormous number of
internal molecular states~\cite{bohn2}. 
 Although we do not assign quantum numbers to the resonant
 states  here, we expect them to be of two basic types, as discussed in~\cite{bohn2}:
 i) coupled- channel shape resonances;  ii) "rotational Feshbach" resonances that change the value of N of one
 or more molecules. This last type of resonance can be extremely long- lived, owing to the difficulty of
  both molecules returning to their rotationless state in a collision. Indeed, one such resonance at $\approx 0.013K$ has a
  width $50 \mu K$.  We will return to this subject in a future publication.
 
Figure 5 shows that the elastic cross section has a very large value near zero energy, corresponding to a scattering
length $a= 270a.u.$ in the present model. 
It is therefore possible that there may be
 an s- wave bound state in the region of negative energies near the threshold of the channel. In this case the cross section
should have $\sim 1/(E+|\varepsilon|)$ dependence~\cite{land} on  energy E and on the energy of the bound state
 $\varepsilon$. 
 
  We have included the dipole- dipole interaction  in these calculations.
However the role of this interaction is very small in general, influencing the cross section at the $1\%$  level and
shifting resonance positions  slightly.
 Likewise, this interaction is only a small perturbation to inelastic scattering.

\subsection{$^{17}O_{2}$- prospects for evaporative cooling.}
For the $|N_{1}N_{2}J_{1}J_{2},M_{J_{1}} M_{J_{2}}> = |0011,11>$ state of interest to trapping experiments, $N_{i}$ and $J_{i}$
are conserved at low energy, since the next energetically available state (with $N=2$) is 11.18K higher in energy~(Fig.1).
Thus the only possible final states are those which differ from the initial one in their projections $M_{J_{1}}$ and $M_{J_{2}}$. 
To accomplish such a transition therefore requires that the angular momentum be carried away in the orbital angular momentum $l$.
Furthermore, the collisions that originate in s- wave channels will be suppressed at energies below the centrifugal barrier of the
d- wave exit channel.
Using an  effective $C_{6}$ coefficient, $C_{6}^{eff}$ from~\cite{bussery} this energy for a partial wave $l$
can be approximated as
\begin{eqnarray}
E_{0}(l)=\frac{\hbar^{2}l(l+1)}{2mr_{b}^{2}}- \frac{C_{6}^{eff}}{r_{b}^{6}},\hspace{2cm}
r_{b}^{2}=\sqrt{\frac{6C_{6}^{eff}m}{\hbar^{2}l(l+1)}}
\end{eqnarray}
For $O_{2}$, $C_{6}^{eff}=80.5a.u.$ and the d-wave threshold energy is $0.013K$.

The main aim of this paper, as previously discussed, is to compare the elastic and  loss rate constants.
 Figure 6a
shows these rates calculated according to~Eqn.(\ref{rate}). 
 Away from  resonances, in the energy range up to $E_{0}\approx 0.013K$ the loss rate constant is indeed strongly
 suppressed.
 A detailed examination of the final states contributing to this loss reveals
  general features which are the similar
to those  for $He- ^{17}O_{2}$ scattering. Namely, elastic scattering, which does not change 
either $M_{J_{1}}$ or $M_{J_{2}}$, is the most
probable result of a collision. The next most likely processes are those for which the final $M_{J}$ differs from the 
initial one by 1 or 2 and
the rates for these processes are smaller than the elastic rate by one -two orders of magnitude. The processes for
which the final $M_{J}$ differs from the initial one by 3 or 4 have rates smaller than for elastic scattering by
 2-4 order of magnitude.
For energies above $\approx 0.013K$  inelastic processes become more probable, with rates only about 7- 10 times smaller than
elastic rates.

The thermally  averaged  elastic and loss rates are relevant to the experimental situation.
If we assume the velocity distribution is Maxwellian characterized by a kinetic temperature T we can calculate
the thermally- averaged rate constant as:
\begin{eqnarray}
\bar{K}(T) =\Bigl(\frac{8k_{B}T}{\pi m}\Bigl)^{1/2} \frac{1}{(k_{b}T)^{2}} \int_{0}^{\infty}E \sigma(E) e^{-E/k_{B}T} dE
\end{eqnarray}
To do this averaging we 
 extrapolate the cross sections to energies $> 1K$ using their values at $E=1K$.

Figure 6b shows these thermally averaged elastic and loss rates.
For the cooling to be efficient the rate of elastic collisions $K_{el}$ must exceed the rate of spin- changing, lossy
collisions $K_{loss}$ by at least two orders of magnitude~\cite{monroe}. 
For $^{17}O_{2}$, 
 below $T \approx 0.01K$ this condition is fullfilled.
However, there is a "relatively dangerous temperature range" 
 above $\approx~0.01K$  where $K_{el}/K_{loss} \approx 7-10$.
 By comparison, consider the equivalent ratio for $He-O_{2}$ scattering, as discussed in Ref.~\cite{bohn2}.
 The fact that $K_{el}/K_{loss}$ is not so large as for $He-O_{2}$ collisions 
originates from the stronger anisotropy and  the deeper PES  for the $O_{2}- O_{2}$ system.
It remains to be seen if the loss rates are sufficiently low to evaporatively cool from buffer- gas temperatures,
$\approx 0.3K$, down to $T<0.01K$, where cooling should be quite efficient.

\subsection{$^{16}O_{2}$}
  It is a different situation for $^{16}O_{2}$ molecules from the point of view of comparing elastic and 
  inelastic cross sections.
The general behavior of the elastic cross section for different channels is similar to that of $^{17}O_{2}$ and
has the same order of magnitude except in the energy region near zero which is very sensitive to the details of the potential
and the reduced mass. With the present PES the $^{16}O_{2}$ scattering length is 28a.u.

 Figure 7 shows the elastic and all the inelastic cross sections for
 the trapped state $|N_{1}N_{2}J_{1}J_{2},M_{J_{1}}M_{J_{2}}> = |1122,22>$. The total number of inelastic channels is 25.
  When the final states are $|1120,20>$ , $|1100,00>$, $|1110,10>$ and so on, 
  i.e when at least one of the molecules changes to the $J=0$ state, the collision is superelastic. 
  It is well known that for
a superelastic channel there is $"\sigma \sim 1/\upsilon"$ threshold law. Thus 
at a low energy there is a substantial loss of molecules from the $|1122,22>$ state.  The same result
 holds for $|1122,11>$ state which is also, of course, susceptible to spin- exchange.
Thus $^{16}O_{2}$ is clearly unstable against collisional losses in a magnetic trap, in sharp contrast to
$^{17}O_{2}$.
 The cross sections for the stretched  states of $^{16}O_{2}$ molecules that 
we are interested in have a  smooth structure in the energy
region up to 1K, implying either a lack of resonances in this region or their large widths.
From Figure 7 we can see just one sharp Feshbach- type resonance near $\approx 0.8K$ belonging to an $l=10$ bound state.
Although there may be other weak resonances, it was not our aim to find all the resonances and identify their nature
in this article.

\section{Conclusion}
In this article we theoretically investigated ground state diatom- diatom collisions in the energy range up to 1K taking 
different isotopomers of oxygen molecules as
a prototype. The main point of our investigation was to estimate 
the ratio of elastic and
inelastic rate constants. The influence of the rotational degrees of freedom is crucial in determining this ratio.
In the case of the odd N- manifold,  it is probably impossible to satisfy the criterion $K_{\rm el}>10^{2}K_{\rm loss}$ 
for stretched states in 
any energy region, because of both the strong anisotropy of the PES and  the existence of
the superelastic channels. In the case of the even N- manifold, namely $N=0$, this criterion can be fullfilled because
for the stretched state $|0011,11>$ there are no superelastic channels.

Even though the required ratio of $K_{\rm el}/K_{\rm loss}$
is not quite met at buffer gas temperatures, it is worth remembering
that the buffer-gas procedure typically produces a far
larger sample of trapped molecules than do the laser cooling
experiments on which evaporative cooling is usually applied.
Thus it is possible that a larger loss rate could be sustained
without harming the overall yield of molecules at ultralow 
temperatures.   Detailed rate-equation simulations of
the cooling process are therefore required, an item to which
we will turn our attention in the future.  

Equally importantly, once the molecules have in fact been cooled
to $\mu$K temperatures, out results imply that the lossy collision
rates have diminished into insignificance, falling to levels 
well below $10^{-14}$ cm$^3$/sec.  This in turn implies that
ultracold spin-polarized $^{17}$O$_2$ gases, like their atomic
counterparts, are experimentally stable and should allow the
production of novel Bose-Einstein condensates.

Beyond the immediate results for our particular model of $^{17}$O$_2$,
the present results have broad implications for many paramagnetic 
molecular species.  Namely, the characteristic suppression of loss
rates below the $d$-wave centrifugal barrier should be a generic
feature for molecules where no superelastic fine-structure-changing
processes exist.  It is also important to assess in detail the influence of
the anisotropy of the PES.  For this purpose further investigations
are necessary.

This work was supported by the National Science Foundation.
\begin{center}
{\large \bf Appendix: The definition of the coefficients.}
\end{center}

Using the dispersion contribution part from~\cite{bussery} and expanding this part in the terms of
spherical reduced harmonics~(\ref{disp}) in the lab-fixed coordinate frame we defined the connection between
our coefficients $d_{L_{A},L_{B},M_{A},M_{B}}$ and coefficients $A, B , C$ from~\cite{bussery}:
\begin{eqnarray}
\nonumber
d_{0,0,0,0}=\frac{1}{3}(2A+8B+8C); d_{2,0,2,0}=2A-4B+2C
\\
\nonumber
d_{2,0,0,0}=d_{0,0,2,0}=\frac{1}{3}(2A+2B-4C)
\\
\nonumber
d_{2,-1,2,1}=d_{2,1,2,-1}=\frac{1}{3}(4A-8B+4C)
\\
\nonumber
d_{2,-2,2,2}=d_{2,2,2,-2}=\frac{1}{3}(2A-4B+2C)
\end{eqnarray}

\begin{figure}
\label{zeeman}
\caption{The lowest-energy Zeeman levels of O$_2$, for odd-$N$ (a) and
even-$N$ (b) rotational manifolds. The weak- field- seeking states of interest here are labeled by their
$M_{J}$ quantum numbers.}
\end{figure}

\begin{figure}
\caption{
The $O_2(^{3}\Sigma_{g}^{-}) - O_2(^{3}\Sigma_{g}^{-} )$ singlet potential for
H- geometry and the contributions from different sets ($L_{A}, L_{B}, L$).
}
\end{figure}

\begin{figure}
\caption{A Sample set of the adiabatic curves for
 $^{16}O_{2}$, in this case for total angular
momentum ${\cal J} = 0$.  In computing these curves only the values
$N=1, 3, 5$ and $l=0,2,4,6$ are included.}
\end{figure}

\begin{figure}
\caption{
Elastic scattering cross sections versus energy for different $l_{max}$ for $^{17}O_{2}$.
For example, $l_{max}=6$ means that $l=0,2,4,6$
partial waves were taken into account. See text for details.
}
\end{figure}

\begin{figure}
\caption{
Elastic partial- wave cross sections  for $^{17}O_{2}$
molecules in their magnetically trapped $|0011,11>$ state.
Odd-l contributions do not exist for this state because of the identical- boson exchange symmetry.
}
\end{figure}

\begin{figure}
\caption{
a)Rate constants versus energy for $^{17}O_{2}-^{17}O_{2}$ collisions with molecules
initially in their $|0011,11>$ state. \\
b)
The thermally averaged elastic and loss rates from 6(a) as a function of temperature.
Elastic collisions strongly dominate spin- changing loss collisions at low temperatures.
$E_{0}$ denotes the energy, in Kelvin units, of the height of the $^{17}O_{2}$ d- wave
centrifugal barrier.
}
\end{figure}

\begin{figure}
\label{112222}
\caption{
Elastic scattering cross sections and all inelastic scattering cross sections
for the initial $|1122,22>$ state of $^{16}O_{2}$. Here channels with $l_{max}=10$,
$N=1$ are included.
}
\end{figure}

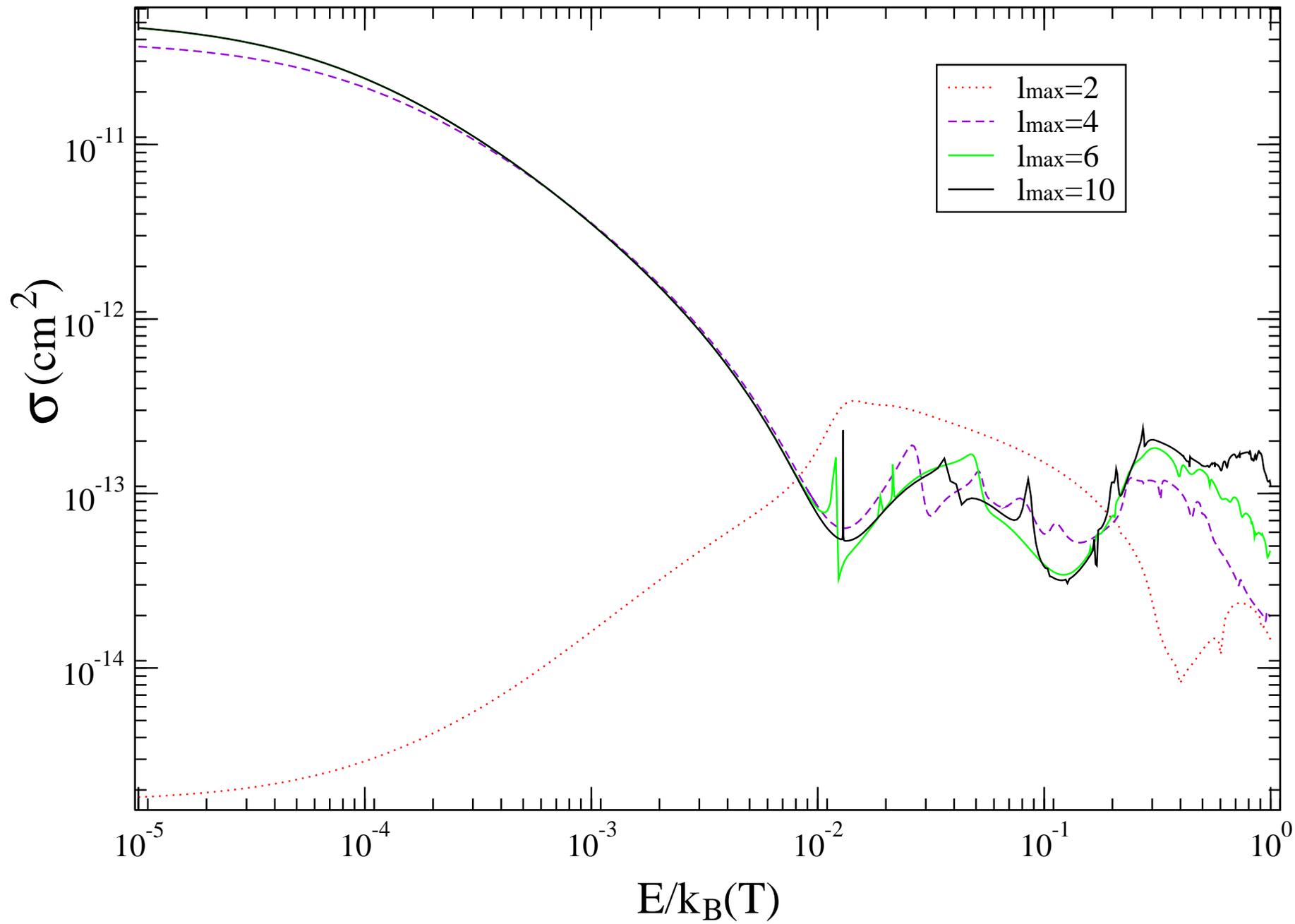
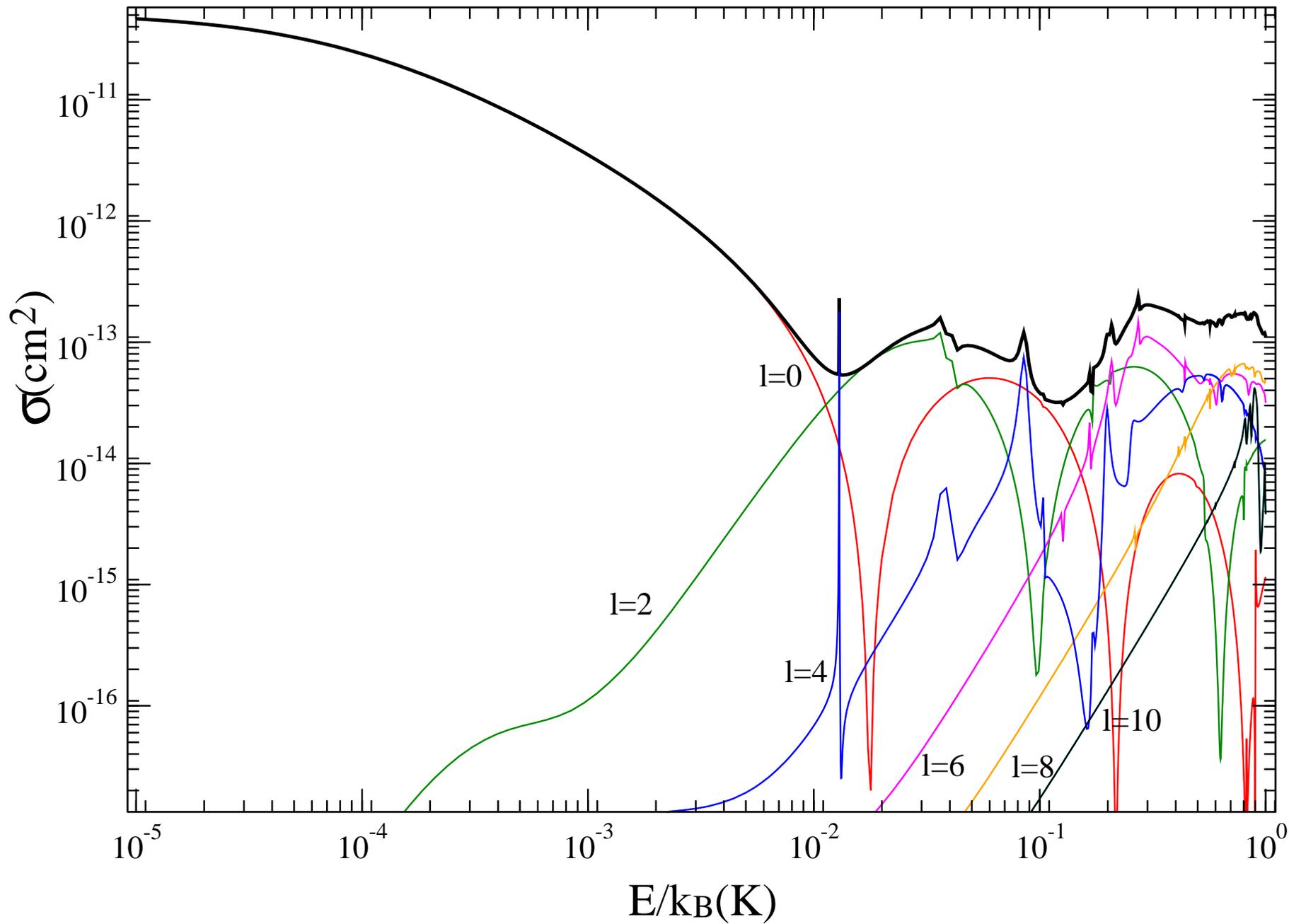
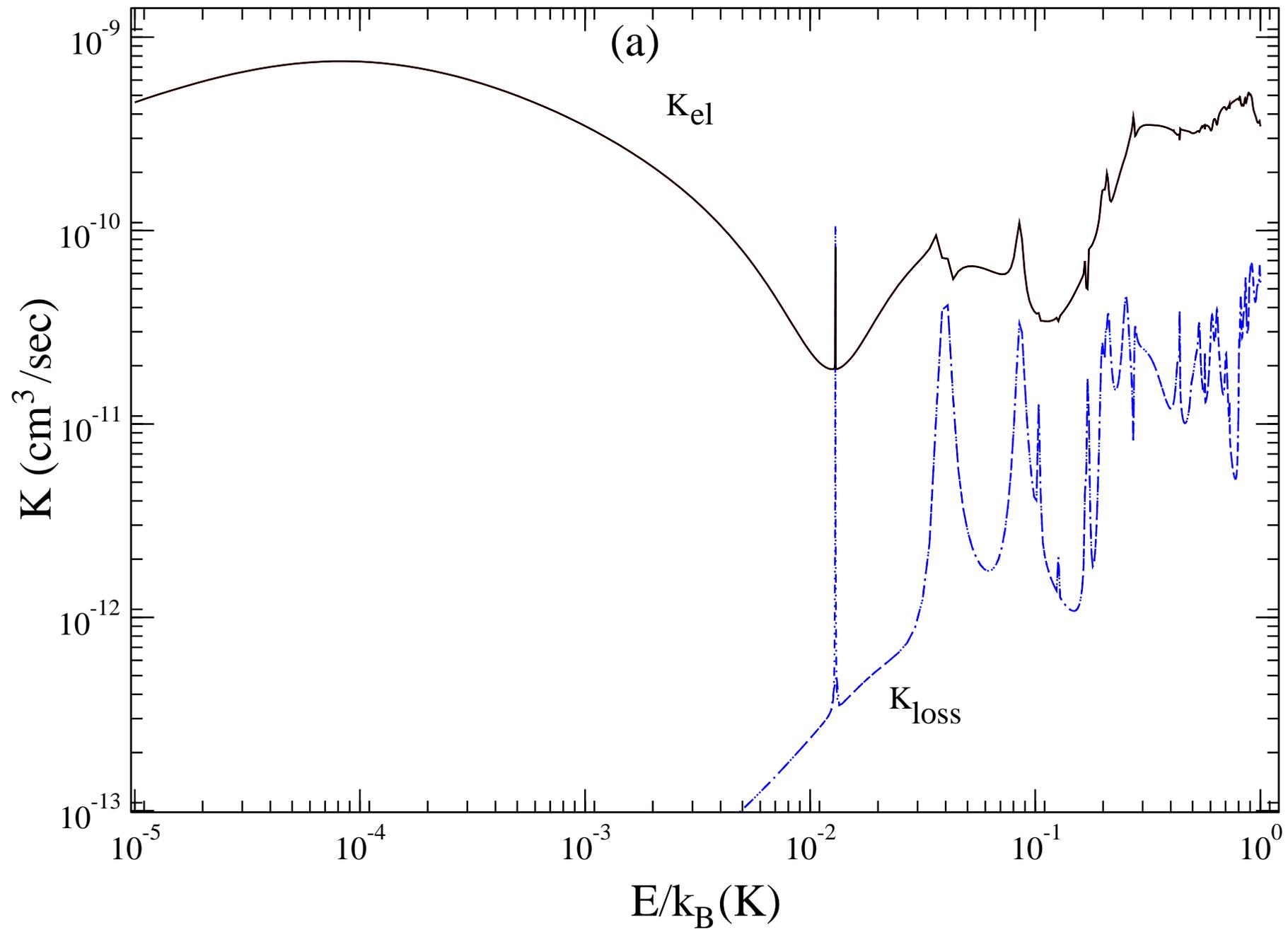
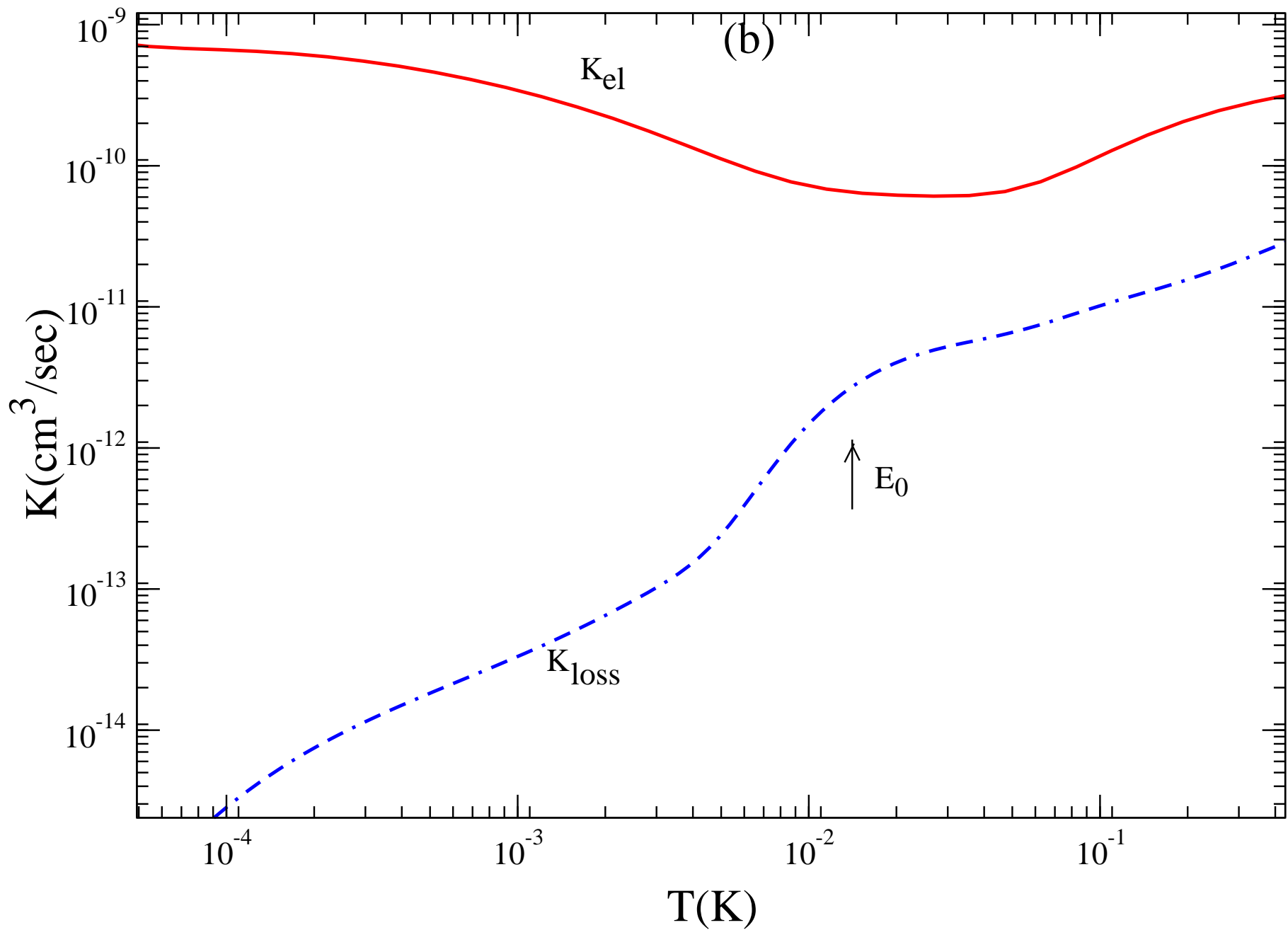
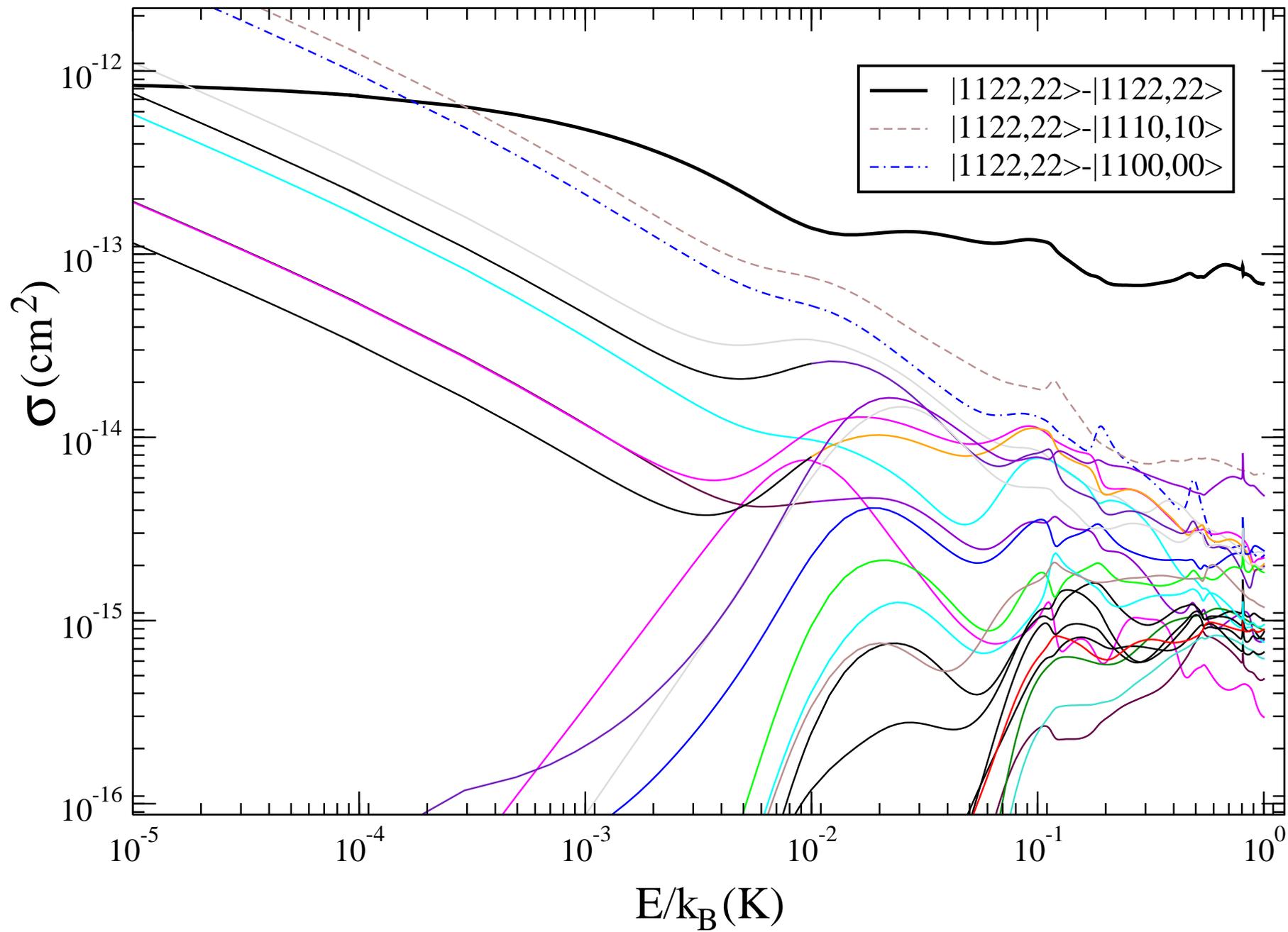
\end{document}